\documentclass[cameraready]{Interspeech}

\title{AdaLTM: Adaptive Layer-wise Task Vector Merging for Categorical Speech Emotion Recognition with ASR Knowledge Integration}

\author[affiliation={1}]{Chia-Yu}{Lee}
\author[affiliation={2}, orcid=0000-0003-2125-5689, equalcontribution]{Huang-Cheng}{Chou}
\author[affiliation={3}, orcid=0009-0001-9800-0193, equalcontribution]{Tzu-Quan}{Lin}
\author[affiliation={4}, orcid=0000-0003-4266-2005, equalcontribution, correspondingauthor]{Yuanchao}{Li}
\author[affiliation={1}]{Ya-Tse}{Wu}
\author[affiliation={2},orcid=0000-0002-1052-6204]{Shrikanth}{Narayanan}
\author[affiliation={1}, orcid=0000-0003-0186-4321, correspondingauthor]{Chi-Chun}{Lee}

\address{
    $^1$ National Tsing Hua University, Hsinchu, Taiwan \\
    $^2$ Signal Analysis and Interpretation Laboratory (SAIL), University of Southern California, USA \\
    $^3$ Graduate Institute of Communication Engineering, National Taiwan University, Taipei, Taiwan \\
    $^4$ University of Edinburgh, Edinburgh, UK
}

\email{aqz7793@gmail.com, yuanchao.li@ed.ac.uk, cclee@ee.nthu.edu.tw}


\newcommand{\err}[1]{{$\pm \scriptstyle #1$}}
\usepackage{comment}
\usepackage[utf8]{inputenc}
\usepackage{textgreek}
\usepackage{graphicx}
\usepackage{subcaption}
\usepackage{tabularx}

\begin{document}

\maketitle

\begin{abstract}
Integrating Automatic Speech Recognition (ASR) into Speech Emotion Recognition (SER) enhances modeling by providing linguistic context. 
However, conventional feature fusion faces performance bottlenecks, and multi-task learning often suffers from optimization conflicts. 
While task vectors and model merging have addressed such conflicts in NLP and CV, their potential in speech tasks remains largely unexplored.
In this work, we propose an Adaptive Layer-wise Task Vector Merging (AdaLTM) framework based on WavLM-Large. 
Instead of joint optimization, we extract task vectors from in-domain ASR and SER models fine-tuned on emotion datasets. 
These vectors are integrated into a frozen base model using layer-wise learnable coefficients. 
This strategy enables depth-aware balancing of linguistic and paralinguistic knowledge across transformer layers without gradient interference.
Experiments on the MSP-Podcast demonstrate that the proposed approach effectively mitigates conflicts between ASR and SER.
\end{abstract}

\section{Introduction and Related Work}
Speech Emotion Recognition (SER) is intrinsically multimodal, relying heavily on both acoustic cues and linguistic content \cite{Lee2005Towarddetectingemotionsin,LeeSpeechEmo-ProcIEEE2023}. 
Consequently, integrating knowledge from Automatic Speech Recognition (ASR) has become a standard paradigm to enhance SER performance \cite{li2022fusing, li2023asr}. 
Early approaches focused on output-level fusion, where textual representations from ASR are combined with acoustic features~\cite{yoon2018multimodal,sahu2019multi}. 
However, this strategy is limited by its sensitivity to ASR transcription errors, particularly on expressive speech~\cite{li2023asr}, and it fails to foster deep, intermediate interactions between modalities~\cite{Li_2024}.

To enable deeper fusion, many have resorted to Multi-Task Learning (MTL), jointly optimizing ASR and SER objectives within a shared acoustic encoder~\cite{cai2021speech,li2022fusing}. 
Yet, this introduces a severe optimization conflict, as the tasks' objectives are fundamentally misaligned. \textit{ASR seeks emotion-invariant representations by suppressing paralinguistic variability, whereas SER relies precisely on such variability to infer emotional states}~\cite{Chou_2024, li2023asr}. 
The resulting gradient interference often degrades the model's ability to learn robust emotional cues, motivating a paradigm shift away from gradient-based joint optimization.

Recently, operating directly in the weight space via \textbf{task vectors} has emerged as a promising, optimization-free alternative~\cite{ilharco2022editing}. 
By defining a task vector $\tau$ as the difference between fine-tuned parameters ($\theta_{\text{ft}}$) and pre-trained parameters ($\theta_{\text{pre}}$), one can algebraically add a task's capability to a base model: $\tau = \theta_{\text{ft}} - \theta_{\text{pre}}$.
While task vectors have been successfully applied across various speech domains~\cite{ramesh2024task, plantinga2024parameter, lin2025speech}, their use in SER remains largely unexplored.

Furthermore, we identify a critical, previously overlooked bottleneck in applying this paradigm to ASR-enhanced SER: \textbf{domain mismatch}. 
We find that simply merging a task vector from a standard, out-of-domain ASR model (e.g., fine-tuned on Librispeech \cite{Panayotov_2015_Librispeech}) yields sub-optimal results. Such models are trained to be emotion-agnostic, actively discarding the rich paralinguistic cues (e.g., laughter, pitch contours) that are vital for SER. 
This aligns with recent findings that highlight the importance of domain consistency and careful integration strategies when leveraging ASR knowledge~\cite{Li_2024,Chou_2024,Li_2025}.

\begin{figure}[!t]
    \centering
\includegraphics[width=\linewidth]{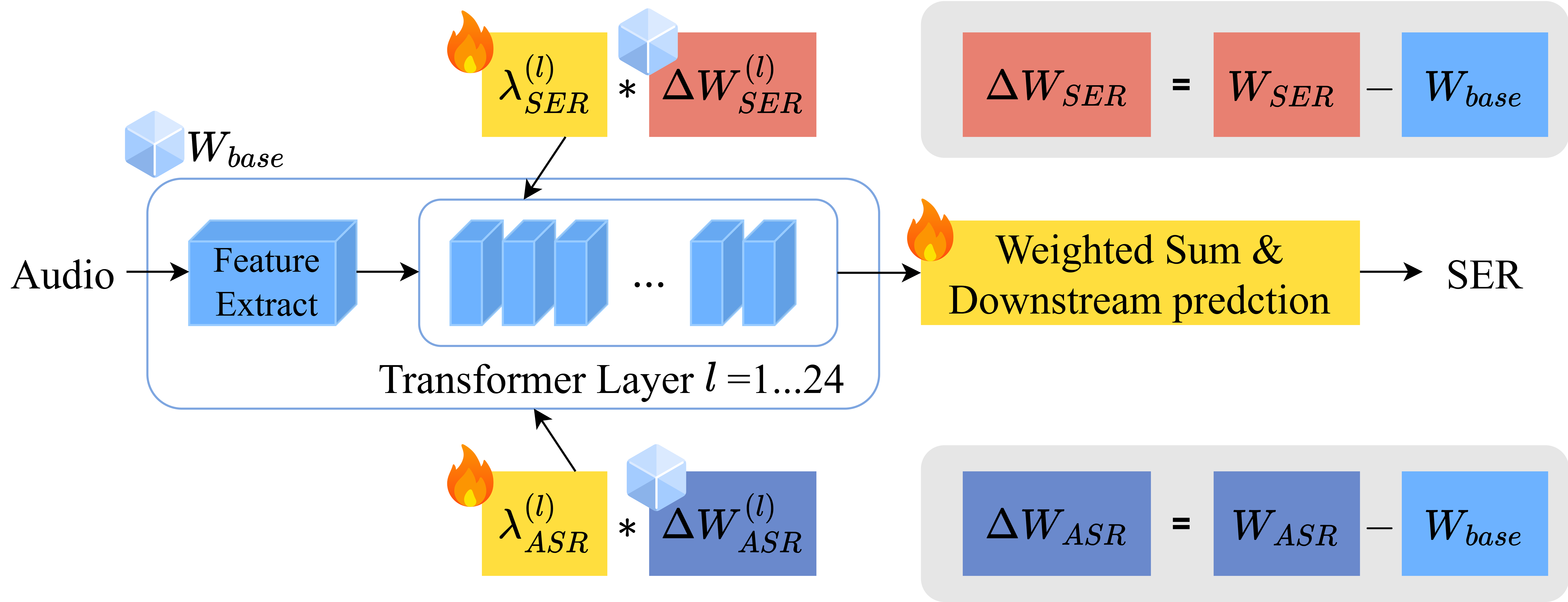}
    \caption{\small The proposed Adaptive Layer-wise Task Vector Merging (AdaLTM) framework. The pre-trained backbone and task vectors ($W_{base}$, $\Delta W_{ASR}$, $\Delta W_{SER}$) remain frozen, while only the layer-wise merging coefficients ($\lambda$) and the downstream prediction head are updated.}
    \label{fig:framework}
    \vspace{-7mm}
\end{figure}

To resolve both the optimization conflict and the domain mismatch, we propose the \textbf{Adaptive Layer-wise Task Vector Merging (AdaLTM)} framework (shown in Figure~\ref{fig:framework}). 
Instead of joint training, we first extract separate task vectors from ASR and SER models that have been independently fine-tuned on the target emotional domain (MSP-Podcast \cite{8003425, busso2025msppodcastcorpus}). 
Building on recent advances in adaptive merging~\cite{yang2024adamerging}, we introduce learnable layer-wise coefficients that dynamically balance and inject these in-domain task vectors into a frozen WavLM-Large backbone \cite{chen2022wavlm}.
We employ WavLM-Large as our backbone, motivated by its leading performance on the SUPERB \cite{yang21c_interspeech} and EMO-SUPERB \cite{Wu_2024_EMO_SUPERB} SER benchmarks.
For each layer $l$, the merged weights $\theta_{\text{merged}}^{(l)}$ are computed as:
\begin{equation}
    \footnotesize
    \theta_{\text{merged}}^{(l)} = \theta_{\text{pre}}^{(l)} + \alpha^{(l)} \tau_{\text{ASR}}^{(l)} + \beta^{(l)} \tau_{\text{SER}}^{(l)},
\end{equation}
where $\alpha^{(l)}$ and $\beta^{(l)}$ are learnable parameters. Our main contributions are threefold:
\begin{itemize}
    \item We introduce a novel adaptive layer-wise model-merging framework (\textbf{AdaLTM}) that integrates ASR knowledge into SER, effectively resolving the optimization conflicts inherent to traditional MTL.\footnote{\scriptsize https://anonymous.4open.science/r/AdaLTM-62A2/}
    \item We establish the critical role of domain consistency in task vector merging, demonstrating that \textit{in-domain} ASR knowledge significantly outperforms out-of-domain alternatives.
    \item We analyze layer-wise merging dynamics and achieve a competitive Unweighted Average Recall of 38.94\% (Macro-F1 scroe of 35.20\%) on the MSP-Podcast dataset.
\end{itemize}

\section{Methodology}
We propose Adaptive Layer-wise Task Vector Merging (AdaLTM), shown in Figure~\ref{fig:framework}, a framework that enhances SER by extracting task-specific knowledge into task vectors and adaptively integrating them into a pre-trained backbone. We employ two distinct layer-wise mechanisms: $\lambda$ for weight-space task vector merging, and $\alpha$ for downstream feature aggregation.

\subsection{Task Vector Formulation}
We employ the WavLM-Large model as our foundational backbone, denoted by its pre-trained weights $W_{base}$. 
To capture domain-specific and task-specific knowledge, we fine-tune $W_{base}$ on distinct target tasks using Differential Learning Rates (DLR). 
Specifically, we fine-tune the base model on the MSP-Podcast dataset to derive both an SER-specific model ($W_{SER}$) and an in-domain ASR model ($W_{ASR}$). 
A task vector represents the direction and magnitude in the weight space required to adapt the base model to a specific task. 
We define the ASR and SER task vectors as the element-wise weight residuals:

\begin{equation}
\footnotesize
\Delta W_{ASR} = W_{ASR} - W_{base}.
\end{equation}
\begin{equation}
\Delta W_{SER} = W_{SER} - W_{base}.
\end{equation}

By isolating these vectors, we capture the transition from generalized acoustic representations to specialized knowledge (i.e., textual mapping for ASR and paralinguistic extraction for SER) without modifying the original backbone.

\vspace{-1mm}
\subsection{Adaptive Layer-wise Merging Strategy}
\vspace{-1mm}
Model merging normally applies a single scaling factor across all layers, which fails to account for the varying levels of abstraction learned at different depths of the transformer. 
To address this, we propose an adaptive Layer-wise merging strategy. 
We partition the WavLM-Large model into 25 distinct layers: one non-encoder layer (comprising the CNN feature extractor and positional embeddings), denoted as index $l = 0$, and the subsequent 24 transformer encoder layers, indexed as $l \in \{1, 2, \dots, 24\}$. 
For each layer $l$, we introduce layer-wise, continuous learnable parameters $\lambda_{ASR}^{(l)}$ and $\lambda_{SER}^{(l)}$ to dynamically scale the respective task vectors. 
The merged weight $W_{merged}^{(l)}$ for the $l$-th layer is formulated as:
\vspace{-1mm}
\begin{equation}
\footnotesize
W_{merged}^{(l)} = W_{base}^{(l)} + \lambda_{ASR}^{(l)} \Delta W_{ASR}^{(l)} + \lambda_{SER}^{(l)} \Delta W_{SER}^{(l)}.
\end{equation}

Both $\lambda_{ASR}^{(l)}$ and $\lambda_{SER}^{(l)}$ are initialized to $0.5$, a value empirically found to provide a stable starting point for the optimization process, ensuring the generalized capabilities of the base model are preserved initially.

\vspace{-1mm}
\subsection{Task-Specific Optimization}
\vspace{-1mm}
During the final phase of emotion training, we utilize the model parameterized by dynamically composed weights $W_{merged}$ as a feature extractor. 
To effectively aggregate the hierarchical features, we extract the hidden states $H^{(l)}$ from all 24 transformer layers and apply a learnable weighted sum mechanism to form the final representation $H_{out}$:

\begin{equation}
\footnotesize
H_{out} = \sum_{l=1}^{24} \alpha_{l} H^{(l)},
\end{equation}

where $\alpha_{l}$ are the normalized, trainable layer weights. 
$H_{out}$ is then fed into a SER Prediction Head for the final classification. 
Crucially, to prevent catastrophic forgetting, the backbone weights ($W_{base}$, $\Delta W_{ASR}$, and $\Delta W_{SER}$) are strictly frozen during this phase. 
The only trainable parameters are the layer-wise merging coefficients $\{\lambda_{ASR}^{(l)}, \lambda_{SER}^{(l)}\}_{l=0}^{24}$, the weighted sum weights $\{\alpha_{l}\}_{l=1}^{24}$, and the parameters of the Emotion Prediction Head. 
This architecture enforces the model to learn how to integrate ASR knowledge for enhancing SER performance.

\vspace{-1mm}
\section{Experimental Setup}
\vspace{-1mm}
\subsection{Dataset and Backbone Models}
\vspace{-1mm}
\label{ssec:dataset_and_models}
All experiments are conducted on the MSP-Podcast (v1.12) corpus~\cite{8003425, busso2025msppodcastcorpus} to ensure domain consistency for both primary SER and auxiliary ASR tasks. 
The SER task is an 8-class classification problem (Anger, Contempt, Disgust, Fear, Happiness, Neutral, Sadness, Surprise). 
To ensure high-quality ASR supervision, we use only samples with human-annotated transcripts, resulting in 89,752 training, 25,232 validation, and 46,366 test samples.

Our framework is built upon the pre-trained WavLM-Large foundation model ($W_{base}$). To extract the necessary task vectors, we employ three fine-tuned model variants:
\begin{itemize}[leftmargin=*,topsep=0pt, partopsep=0pt, itemsep=2pt]
    \item \textbf{Primary SER Model ($W_{SER}$):} A WavLM-Large model fine-tuned on MSP-Podcast for SER~\cite{feng2025voxprofilespeechfoundationmodel}\footnote{\scriptsize https://huggingface.co/tiantiaf/wavlm-large-categorical-emotion}, establishing our acoustic baseline with a MaF1 of 35.56\%.
    \item \textbf{In-domain ASR Model ($W_{ASR_{in}}$):} A WavLM-Large model fine-tuned on MSP-Podcast transcripts, achieving a robust WER of 23.09\% on the test set and providing domain-aligned linguistic knowledge.
    \item \textbf{Out-of-domain ASR Model ($W_{ASR_{out}}$):} A standard WavLM-Large model fine-tuned on LibriSpeech 100h\footnote{\scriptsize https://huggingface.co/patrickvonplaten/wavlm-libri-clean-100h-large}, used as a baseline to demonstrate the importance of domain consistency. It yields a much higher WER of 37.86\%.
\end{itemize}

\begin{table}[!t]
  \centering
  \fontsize{7}{9}\selectfont 
  \setlength{\tabcolsep}{1.2pt} 
  \renewcommand{\tabularxcolumn}[1]{m{#1}} 
  \caption{\small Performance comparison of different ASR integration strategies and merging configurations on the MSP-Podcast dataset. All metrics are reported in (\%). Best results for our proposed paradigm are in \textbf{bold}. \textbf{Pre.}: Precision; \textbf{MaF1}: Macro-F1. We report the 95\% confidence interval (CI) for each SER result using the toolkit~\cite{Confidence_Intervals}.}
  \vspace{-3mm}
  \label{tab:unified_results}
  \begin{tabularx}{\columnwidth}{@{}Xcccc@{}}
    \toprule
    \textbf{Method/Setup} & \textbf{UAR} & \textbf{Pre.} & \textbf{MaF1} & \textbf{WER} \\
    \midrule
    \multicolumn{5}{c}{\textit{\textbf{Part 1: Multi-Task Learning (MTL) Baselines}}} \\
    \midrule
    WavLM-Large (Fully Trainable) & 29.54\err{0.38}& 33.35\err{1.47} & 28.40\err{0.51} & 99.12 \\
    MTL w/ static init. ($W_{base} + 0.5\Delta W$) & 29.21\err{0.38} & 38.30\err{2.46} & 29.06\err{0.50} & 66.37 \\
    \midrule
    \multicolumn{5}{c}{\textit{\textbf{Part 2: Ablation Study of Our Merging Paradigm (AdaLTM)}}} \\
    \midrule
    Setup 1: Baseline (Frozen Backbone) & 37.05\err{0.67} & 34.46\err{0.42} & 34.46\err{0.47} & - \\
    Setup 2: ASR-Only Vector & 37.57\err{0.67} & 34.43\err{0.38} & 33.56\err{0.42} & - \\
    Setup 3: SER-Only Vector & \textbf{39.09}\err{0.60} & \textbf{34.80}\err{0.39} & \textbf{35.41}\err{0.44} & - \\
    Setup 4: Proposed Dual-Vector & 38.94\err{0.61} & 34.26\err{0.42} & 35.20\err{0.48} & - \\
    \midrule
    \multicolumn{5}{c}{\textit{\textbf{Part 3: Importance of Domain (Dual-Vector Merging)}}} \\
    \midrule
    Out-of-Domain Dual-Vectors & 38.68\err{0.63} & 34.20\err{0.43} & 34.84\err{0.48} & - \\
    In-Domain Dual-Vectors (\textbf{Ours}) & \textbf{38.94}\err{0.61} & \textbf{34.26}\err{0.42} & \textbf{35.20}\err{0.48} & - \\
    \midrule
    \multicolumn{5}{c}{\textit{\textbf{Part 4: Importance of Granularity (Merging Strategy)}}} \\
    \midrule
    Static Global Merging ($\lambda=0.5$) & 38.30\err{0.61} & \textbf{34.71}\err{0.46} & \textbf{35.73}\err{0.51} & - \\
    Adaptive Global Merging & 38.93\err{0.60} & 34.02\err{0.43} & 34.85\err{0.47} & - \\
    Adaptive Layer-wise Merging (\textbf{Ours}) & \textbf{38.94}\err{0.61} & 34.26\err{0.42} & 35.20\err{0.48} & - \\
    \bottomrule
  \end{tabularx}%
  \vspace{-6mm}
\end{table}

\vspace{-1mm}
\subsection{Comparison Setups}
\label{ssec:comparison_setups}
\vspace{-1mm}
We design two sets of experiments to validate our approach. 
First, to demonstrate the synergy of our dual-vector merging, we conduct a comprehensive ablation study, with results presented in Part 2 of Table~\ref{tab:unified_results}. 
The setups include: \textbf{(1) Baseline}, a frozen WavLM backbone without merging; \textbf{(2) ASR-Only}, merging only the in-domain ASR vector; \textbf{(3) SER-Only}, merging only the SER vector; and \textbf{(4) Proposed Dual-Vector}, our complete framework integrating both.

Second, to justify the necessity of layer-wise granularity, we compare our proposed \textbf{Adaptive Layer-wise} strategy against two global baselines (Part 4 of Table~\ref{tab:unified_results}): a \textbf{Static Global} merge with a fixed $\lambda=0.5$ for all layers, and an \textbf{Adaptive Global} merge that learns a single shared $\lambda$ across all layers.

\subsection{Implementation Details and Metrics}
\label{ssec:implementation}
To prevent catastrophic forgetting, the pre-trained backbone ($W_{base}$) and task vectors ($\Delta W$) are frozen during all downstream experiments. 
Training is restricted to the layer-wise merging coefficients $\lambda$ (initialized to 0.5), the weighted sum weights $\alpha_l$, and the emotion prediction head.
Each task vector introduces only 25 additional learnable parameters: 24 layer-wise $\lambda$ coefficients for the transformer layers and 1 coefficient for the audio front-end block.
These parameters are optimized using the AdamW optimizer \cite{loshchilov2018decoupled} with a learning rate of $1.0 \times 10^{-4}$ and a batch size of 32. 
To address class imbalance in the MSP-Podcast dataset, we employ a class-balanced soft cross-entropy loss~\cite{Cui_2019_CVPR, chou23_interspeech}. 
For each experiment, the model checkpoint with the lowest validation loss from 100 training epochs is selected for evaluation. 
Performance is primarily evaluated using Unweighted Average Recall (UAR), supplemented by Precision and Macro-F1 (MaF1) for a comprehensive analysis. 
All experiments were conducted using the PyTorch framework \cite{NEURIPS2019_bdbca288} on two NVIDIA V100 GPU (64GB).

\begin{figure}[!t]
    \centering
    \includegraphics[width=.7\linewidth]{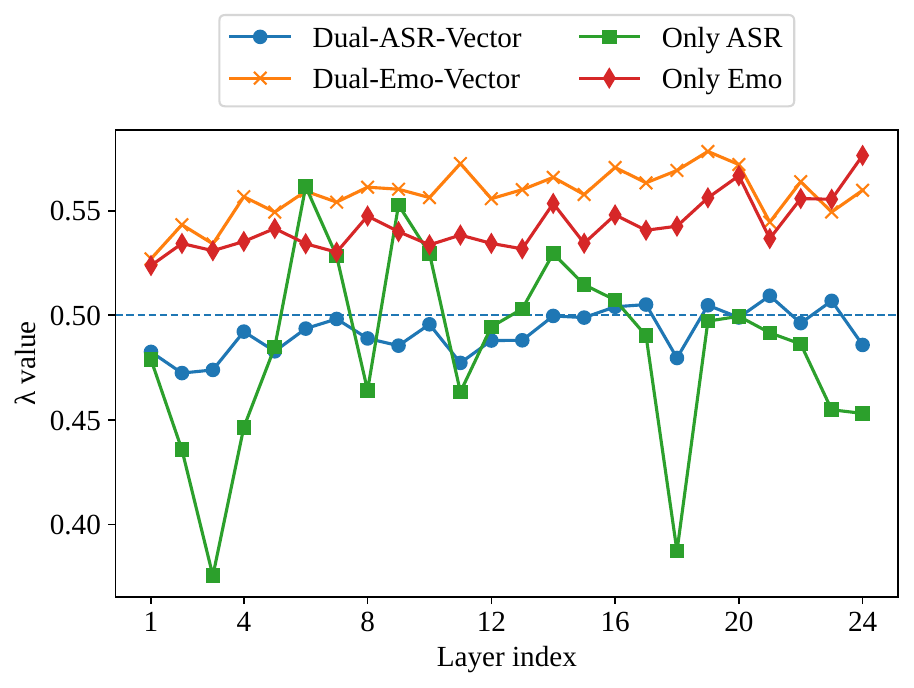}
    \vspace{-3mm}
    \caption{\small Layer-wise Dynamics: Dual vs. Single Task Vectors. \textcolor{blue}{Blue line}: Proposed dual-vector ASR. \textcolor{orange}{Orange line}: Dual-vector SER. \textcolor{teal}{Green line}: Only-ASR setup. \textcolor{red}{Red line}: Only-SER setup.}
    \label{fig:dual_vector}
    \vspace{-4mm}
\end{figure}

\vspace{-1mm}
\section{Results and Analyses}
\vspace{-1mm}
\subsection{Baseline Comparison: Overcoming Conflicts}
\label{ssec:baseline_comparison}
\vspace{-1mm}
To evaluate the limitations of conventional MTL, we compare our framework against two fully trainable baselines updating via joint ASR and SER losses. 
To mitigate initialization bias, the second baseline initializes with statically merged task vectors ($W_{base} + 0.5 \Delta W_{ASR} + 0.5 \Delta W_{SER}$).

As shown in Part 1 of Table~\ref{tab:unified_results}, conventional MTL approaches exhibit severe gradient interference, or the ``seesaw effect.'' 
Although static initialization constrains auxiliary WER to 66.37\% (vs. 99.12\% for the vanilla backbone), the primary SER performance inevitably collapses to a UAR of 29.62\%. This confirms that jointly optimizing a single backbone for diametrically opposed tasks: emotion-invariant ASR versus emotion-rich SER, degrades acoustic representations. 

In contrast, AdaLTM eliminates these conflicts by strictly freezing the backbone during adaptive layer-wise merging. 
This strategy achieves a UAR of 38.62\% and Precision of 34.55\%, an absolute improvement of over 8.4\% compared to the best MTL epoch, substantiating that multi-task knowledge is most effectively unified via weight-space merging rather than joint backpropagation.
Furthermore, compared to the fully trainable MTL baseline that updates 311.7M parameters, accounting for 98.6684\% of the 315.9M total model parameters, AdaLTM updates only 0.46M parameters, corresponding to just 0.1463\% of the logical total model size.

\begin{figure}[!t]
    \centering
    \includegraphics[width=.7\linewidth]{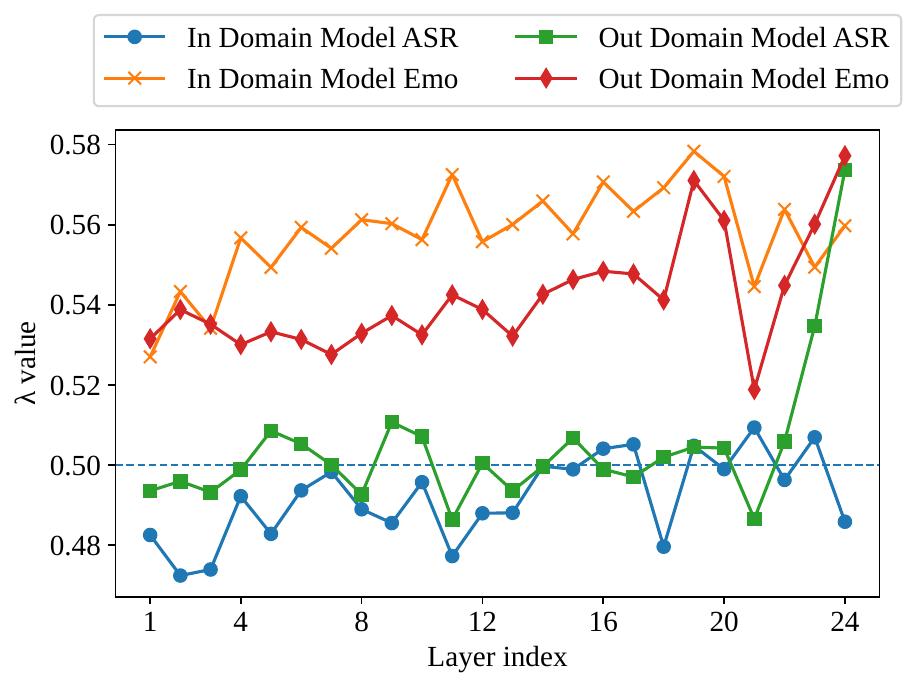}
    \vspace{-3mm}
    \caption{\small Impact of Domain Consistency on Task Vector Merging. \textcolor{blue}{Blue line}: In-domain ASR vector. \textcolor{orange}{Orange line}: In-domain SER vector. \textcolor{teal}{Green line}: Out-domain ASR vector. \textcolor{red}{Red line}: Out-domain SER vector.}
    \label{fig:in_out_domain}
    \vspace{-7mm}
\end{figure}

\subsection{The Synergy of Complementary Task Vectors}
Part 2 of Table~\ref{tab:unified_results} presents a comprehensive ablation study validating the necessity of our dual-vector approach. 
The progression of these results clearly illustrates the additive and synergistic value of merging complementary task vectors. 
The progression of these results clearly illustrates the additive and synergistic value of merging complementary task vectors.

The experiment begins with \textbf{Setup 1 (Baseline)}, which uses the frozen WavLM backbone as a static feature extractor. This yields a UAR of 37.05\%, reflecting the performance ceiling of relying solely on generalized, pre-trained acoustic representations for this task. 

Next, \textbf{Setup 2 (ASR-Only)} introduces linguistic knowledge by merging exclusively with the in-domain ASR task vector, raising the UAR to 37.57\%. 
This demonstrates that conversational linguistic context provides a fundamental level of emotional discriminability, but the marginal improvement suggests it lacks the explicit paralinguistic fine-tuning required for robust SER.

Conversely, \textbf{Setup 3 (SER-Only)} captures these crucial paralinguistic nuances by merging only with the SER task vector. 
This achieves a much stronger UAR of 39.09\%, establishing a rigorous single-vector merging baseline. 
The absolute gain of 2.04\% UAR over the baseline proves that adaptively scaling a frozen, task-specific residual vector is a vastly superior strategy to using the base model's raw features.

Finally, \textbf{Setup 4 (Proposed Dual-Vector)} integrates both ASR and SER vectors simultaneously, achieving a highly competitive peak UAR of 38.94\%. 
This result demonstrates a definitive synergistic effect. 
By independently scaling the ASR and SER vectors, our adaptive mechanism seamlessly interlocks textual semantics (``what is being said,'' from Setup 2) with acoustic prosody (``how it is spoken,'' from Setup 3). 

The marginal 0.15\% UAR reduction compared to the SER-Only setup reflects the physical constraints of representational crowding: accommodating both emotion-invariant lexical mappings and emotion-rich prosody within a fixed-capacity model introduces slight parameter competition. 
This observation is highly consistent with recent findings in adaptive model merging \cite{yang2024adamerging}, which demonstrate that individual task-specific expert models inherently establish a performance upper bound, making minor degradations during multi-task weight fusion an expected theoretical outcome.

However, the critical takeaway is the framework's robustness. 
Unlike joint-training approaches that suffer from catastrophic interference, our layer-wise mechanism successfully navigates this crowding, preserving the expert-level capability of the SER vector while safely integrating textual semantics. 
This combined integration definitively outperforms the standard static baseline (Setup 1) by a 1.89\% absolute margin in UAR.

\subsection{Impact of Domain and Layer-wise Knowledge}
Our experiments also underscore the critical importance of both domain alignment and merging granularity, with results detailed in Part 3 and 4 of Table~\ref{tab:unified_results}.

\textbf{Domain Consistency is Key:} When replacing the in-domain ASR vector with an out-of-domain (LibriSpeech-tuned) one, the UAR drops from 38.94\% to 38.68\%. 
While this performance is still strong, the degradation confirms that domain-aligned linguistic features provide a more effective semantic anchor, as visually supported by our layer-wise analysis in Section~\ref{ssec:layer_wise_dynamics}.

\textbf{Layer-wise Granularity Matters:} We compared our proposed adaptive layer-wise merging against two global strategies. A static global merge with a fixed $\lambda=0.5$ yields a UAR of 38.30\%. 
An adaptive global strategy, which learns a single shared $\lambda$ for all layers, improves this to 38.93\%. 
However, our adaptive layer-wise approach achieves the highest UAR of 38.94\%, demonstrating that providing the model with the flexibility to balance ASR and SER knowledge differently across the network's depth is crucial for resolving representational conflicts and achieving optimal performance.
\subsection{Layer-wise Dynamics: Unveiling Multi-Task Synergy and Domain Mismatch}
\label{ssec:layer_wise_dynamics}
To physically interpret our merging results, Figure~\ref{fig:dual_vector} visualizes the learned layer-wise $\lambda$ trajectories across the 24 transformer layers. 
These distributions reveal exactly how the model manages representational crowding and domain interference.

\textbf{In-Domain Synergy:} Linguistic Anchoring and Prosodic Dominance.
In our proposed dual-vector setup, the in-domain ASR weights ($\lambda_{ASR_{in}}$, blue line) stabilize remarkably near the 0.5 baseline, sharply contrasting the severe volatility seen when the ASR vector is used alone (green line). 
This proves that when acoustic prosody is available, the ASR vector no longer struggles to predict emotions; instead, it functions as a stable linguistic anchor. 
Supported by this semantic foundation, the SER vector achieves absolute prosodic dominance. 
Across the middle-to-deep layers, the dual-setup SER weights (orange) track consistently higher than the single-setup SER weights (red). 
The adaptive mechanism confidently amplifies paralinguistic features, physicalizing the $1+1 > 1$ synergy observed in our UAR metrics.

\textbf{Out-Domain Mismatch:} Feature Suppression and Optimization Chaos.
Conversely, introducing an out-of-domain task vector severely disrupts this balance. 
First, the in-domain SER weights ($\lambda_{Emo_{out}}$) are actively suppressed across the middle and deep layers, demonstrating that emotion-agnostic textual features directly hinder the extraction of paralinguistic cues. 
Furthermore, the out-domain ASR weights ($\lambda_{ASR_{out}}$) exhibit violent fluctuations, culminating in an unnatural, chaotic spike in the final semantic blocks (layers 20–24). 
This optimization chaos occurs because the model erratically up-scales conflicting, rigid textual mappings in a failed attempt to minimize the loss. 
This visually confirms that without strict domain alignment, adaptive merging degrades into gradient instability and representational interference.

\vspace{-2mm}
\section{Discussion and Conclusion}
\vspace{-1mm}
This work demonstrates that operating in the weight space via Adaptive Layer-wise Task Vector Merging (AdaLTM) provides an effective alternative to conventional multi-task learning for integrating auxiliary ASR knowledge into SER. By avoiding joint backpropagation, the proposed framework eliminates gradient interference between ASR and SER objectives, which commonly limits traditional joint training paradigms. Our layer-wise analysis further suggests that deeper transformer blocks benefit more from domain-aligned linguistic representations, supporting the importance of both domain consistency and layer-wise granularity in multi-task model merging.

Extensive experiments confirm that domain alignment plays a critical role in successful task vector integration. Incorporating an in-domain ASR task vector (MSP-Podcast) consistently improves performance, particularly for high-arousal emotions characterized by strong acoustic variability, whereas merging an out-of-domain vector (LibriSpeech) leads to performance degradation. These findings highlight that auxiliary linguistic knowledge must be both structurally and distributionally compatible with the target emotional domain. Overall, the proposed approach achieves a UAR of 38.62\%, demonstrating that task vector merging is viable for ASR-enhanced SER.

Despite these promising results, several \textbf{limitations} remain. First, the effectiveness of AdaLTM depends on the availability of in-domain transcriptions to fine-tune the auxiliary ASR model. For under-resourced emotional datasets lacking reliable transcripts, extracting a highly compatible ASR task vector remains challenging. Second, although the downstream adaptation stage is parameter-efficient (updating only the layer-wise coefficients ($\lambda$) and the prediction head), the initial extraction of task vectors requires fine-tuning separate foundation models, introducing additional computational overhead. \textbf{Future work} will focus on resolving these limitations toward more generalizable and zero-shot SER scenarios.

\section{Acknowledgments}
We acknowledge computing resources from the National Center for High-Performance Computing (NIAR, Taiwan). Supported by NSTC, Taiwan (112-2628-E-007-012-MY3; 114-2917-I-564-030 to Huang-Cheng Chou), NSF (IIS 2311676), and ODNI IARPA ARTS (D2023-2308110001). Thanks to Tiantian Feng and Hung-yi Lee for valuable feedback.


\section{Generative AI Use Disclosure}
Generative AI tools were used only for minor improvements in language and presentation. 
No AI system was used to generate, modify, or interpret the scientific content of this manuscript. 
All authors are fully accountable for the originality and validity of the research.

\bibliographystyle{IEEEtran}
\bibliography{mybib}

\end{document}